\documentclass[12pt]{article}
\usepackage{epsfig}
\begin{document}
\setlength{\baselineskip}{0.30in}
\newcommand{\nc}{\newcommand}
\newcommand{\beq}{\begin{equation}}
\newcommand{\eeq}{\end{equation}}
\newcommand{\be}{\begin{eqnarray}}
\newcommand{\ee}{\end{eqnarray}}
\newcommand{\num}{\nu_\mu}
\newcommand{\nue}{\nu_e}
\newcommand{\nut}{\nu_\tau}
\newcommand{\nus}{\nu_s}
\newcommand{\mnus}{M_s}
\newcommand{\taus}{\tau_{\nu_s}}
\newcommand{\nnt}{n_{\nu_\tau}}
\newcommand{\rnt}{\rho_{\nu_\tau}}
\newcommand{\mnt}{m_{\nu_\tau}}
\newcommand{\tnt}{\tau_{\nu_\tau}}
\newcommand{\bi}{\bibitem}
\newcommand{\rar}{\rightarrow}
\newcommand{\lar}{\leftarrow}
\newcommand{\lrar}{\leftrightarrow}
\newcommand{\dm}{\delta m^2}
\newcommand{\so}{\, \mbox{sin}\Omega}
\newcommand{\co}{\, \mbox{cos}\Omega}
\newcommand{\sotil}{\, \mbox{sin}\tilde\Omega}
\newcommand{\cotil}{\, \mbox{cos}\tilde\Omega}
\makeatletter
\def\alt{\mathrel{\mathpalette\vereq<}}
\def\vereq#1#2{\lower3pt\vbox{\baselineskip1.5pt \lineskip1.5pt
\ialign{$\m@th#1\hfill##\hfil$\crcr#2\crcr\sim\crcr}}}
\def\agt{\mathrel{\mathpalette\vereq>}}

\newcommand{\eq}{{\rm eq}}
\newcommand{\tot}{{\rm tot}}
\newcommand{\M}{{\rm M}}
\newcommand{\coll}{{\rm coll}}
\newcommand{\ann}{{\rm ann}}
\makeatother

\begin{center}
\vglue .06in
{\Large \bf { PHASE TRANSITIONS DURING INFLATION AND CHEMICALLY 
INHOMOGENEOUS UNIVERSE}}
\bigskip
\\{\bf A.D. Dolgov
\footnote{Also: ITEP, Bol. Cheremushkinskaya 25, Moscow 113259, Russia.}
 \\[.05in]
{\it{INFN section of Ferrara\\
Via del Paradiso 12,
44100 Ferrara, Italy}
}}
\\[.40in]
\end{center}

\begin{abstract}
Several models of baryo(lepto)- genesis that give rise to large 
inhomogeneities in the composition of the universe are presented. 
In particular, a variation of primordial abundances of light elements 
by the factor 2-5 at large distances is predicted. A cosmological 
model of baryonic island is considered. Creation of domains with very 
large baryon and antibaryon number is described. Such domains mostly 
collapsed into primordial black holes that could be the dominant part 
of (non-standard) cosmological cold dark matter with a widely dispersed 
mass spectrum. A non-collapsed part of these bubbles might make clouds 
of matter or antimatter with an enriched abundances of metals. A 
mechanism for creation of such exotic objects can be realized by
mixed order phase transitions induced by the inflaton field.
\end{abstract}

\section{Introduction \label{intr}}

It is commonly believed that the universe is the same everywhere (at
least inside the present day horizon). This is the well known
Copernicus or cosmological principle, or the principle of cosmic 
democracy. Observational astronomical data quite well agree with 
this principle and especially strong argument comes from the perfect 
angular smoothness of the cosmic microwave background radiation (CMBR).
However, chemical inhomogeneities may escape detection by CMBR either
because they do not necessarily imply inhomogeneities in mass or energy 
densities or if their size is too small to be observable by the
contemporary CMBR telescopes. A detection of these objects by other 
astronomical methods may be inhibited by their large distance from the 
Earth. Thus at the present level of our knowledge, there seems 
to be plenty of room for ``pieces'' of universe with an exotic 
matter content. Of course the hypothesis of cosmologically large 
chemical inhomogeneities looks quite striking but it already existed 
in different forms for quite a long time (e.g. the idea of domains 
of cosmic antimatter). Several simple theoretical 
models predict such phenomena. The universe with non-homogeneous 
chemistry can be created if certain phase transitions occurred at 
inflationary stage leaving behind astronomically large bubbles with 
different physical conditions. Such phase transitions should have a 
rather unusual character. The effective mass of a scalar (Higgs-like) 
field might be negative but only during a finite (and rather small) 
period of time. This period should take place not too long before 
the end of inflation. These properties could be realized by a simple 
coupling of this Higgs-like field (order parameter) to the inflaton. 
The models with this property predict formation of chemically different
domains in the early universe that might be astronomically large but 
not too large and sufficiently rare to escape existing 
observational bounds.

Since this is the last school on cosmological phase transitions of this
century and even of the Second Millennium it is tempting to speculate
about cosmology in the coming XXI century. To verify or to disprove
Copernicus principle seems to be a great challenge for the future 
(hopefully not too distant one).

The content of the lecture is the following. First, we discuss the model
of baryo(lepto)-genesis that could give rise to bubbles of chemically
different phases in the universe. In sec. 3 we present a model of 
inhomogeneous lepto-genesis that predicts an existence of
cosmologically large regions with strongly varying 
primordial abundances of light elements. The simplest version of the 
model predicts 2/3 of the sky with the so called normal abundances 
(25\% of $He^4$ and about $3\cdot 10^{-5}$ of deuterium), 
1/6 of the sky with much richer fraction of light elements 
(50\% of $He^4$ and about $15\cdot 10^{-5}$ of $D$), and 
1/6 of the sky with a poor production of light elements 
(12\% of $He^4$ and about $1.5\cdot 10^{-5}$ of $D$). In the next
section an inhomogeneous baryogenesis leading to formation of baryonic
and anti-baryonic cosmological islands is discussed.
After that we consider a possible mechanisms of creation
of cosmic antimatter in the universe 
dominated by baryons. Related to that is a model of formation of 
primordial black holes that could deliver a rather unusual form of 
cold dark matter with a very broad mass spectrum of constituent 
``particles''.

\section{Baryo- and Lepto- Genesis \label{bar}}

After Sakharov~\cite{sakharov67} formulated three general principles 
of baryogenesis: non-conservation of baryonic charge,
breaking of particle-antiparticle symmetry, and
deviation from thermal equilibrium,
a variety of models were invented that predict, in particular, noticeable
inhomogeneities in baryon or lepton number densities (isocurvature 
perturbations) which might lead to chemically inhomogeneous universe
(for a review see ref.~\cite{dolgov92}). We will describe here one of the
models that is especially interesting from this point of view, namely
the Affleck and Dine scenario~\cite{affleck85}. The model is based on
a supersymmetric theory that possesses the following generic properties.
First, there exist scalar fields with non-zero baryonic and/or leptonic
charges. We denote these scalar baryons or leptons generically as 
$\chi$. Such fields are obligatory in any supersymmetric model. For 
example they could be super-partners of quarks and leptons. Second, 
the potential of these fields $\chi$ have the so called flat directions
along which the energy does not rise. Such flat direction might exist, 
in particular, prior to spontaneous symmetry breaking before particle 
masses were generated by the Higgs mechanism. 

During inflation scalar fields (in contrast to fermions) may form a 
condensate by rising quantum fluctuations along flat 
directions~\cite{bunch78,vilenkin82,linde82,starobisky82}. 
A large baryonic charge can be stored in this condensate if baryon 
number is not conserved. The subsequent decay of this condensate in 
the course of processes that conserve baryonic charge would 
create a cosmological baryon (or lepton) asymmetry that might reach very 
large values, even close to unity. Recall that the observed value is 
\be
\beta = N_B /N_\gamma \approx (3-5)\cdot 10^{-10}
\label{beta}
\ee
The model permits to create simultaneously a small baryon 
asymmetry in agreement with observations together with a large lepton
asymmetry~\cite{dolgov91,dolgov92,casas99}. Moreover the asymmetries may
be inhomogeneous and the scale of variation of lepton asymmetry may be
much smaller then the scale of variation of baryon 
asymmetry~\cite{dolgov91,dolgov92}.

To illustrate the main features of this scenario let us consider the
toy model with the following potential term:
\be
U(\chi) = m^2 |\chi |^2 + \lambda_1  |\chi |^4 + 
\lambda_2 \left( \chi^4 + \chi* ^4\right)
\label{uchi}
\ee
In the FRW cosmology, where the metric is given by 
$ds^2 = dt^2 -a^2 (t) dr^2$, the equation of motion for $\chi$ has the
form:
\be
\left( \partial^2_t - \partial_k^2 /a^2 \right) \chi +
3H \dot\chi +U'(\chi) = 0
\label{eqmot}
\ee
where $H=\dot a/a$ is the Hubble parameter.

If $\lambda_2 = \lambda_1/2$ the potential has flat directions
in the limit $m=0$ and the field could easily evolve along them.
It is convenient to introduce phase and absolute value of the field:
\be
\chi = r e^{i\theta}
\label{rtheta}
\ee
and in terms of this quantities flat directions go along 
$1+\cos 4\theta =0$. 

The field $\chi$ may possess leptonic or baryonic charge (or a mixture
of both) and the corresponding vector current is given by
\be
j_\mu = i\left( \chi^* \partial_\mu \chi -\partial_\mu \chi^* \chi
\right)
\label{jmu}
\ee
According to the equation of motion (\ref{eqmot}) the current is not
conserved if $\lambda_2 \neq 0$:
\be
\partial_\mu j^\mu = 2i\lambda_2 \left[ \chi^4 - (\chi^*)^4\right] 
\label{dj}
\ee
In the homogeneous case the equation of motion takes the form:
\be
\ddot \chi + 3H\dot \chi +U' (\chi) =0  
\label{ddotchi}
\ee
It is the Newtonian equation of motion for a point-like body in two 
dimensional space of the complex plane $\chi$ in the potential 
$U (\chi)$. Here $\ddot \chi$ is the acceleration, $U'$ is the
force, and $3H\dot\chi$ is the ``liquid'' friction term. Baryonic charge
density is given by:
\be
B(t) \equiv j_0 = r^2 \dot\theta
\label{bt}
\ee
and, in terms of the mechanical analogy, coincides with the angular 
momentum of the body moving in the potential $U$. If the potential is 
spherically symmetric, i.e. it depends only on $r=|\chi|$ the angular
momentum (or baryonic charge) is conserved. The presence of valleys
breaks the symmetry and accordingly the conservation of $j_0$. 

During inflation, when/if $m \ll H$, the field $\chi$ is infrared unstable
and its quantum fluctuations rise in accordance with the 
law~\cite{vilenkin82,linde82,starobisky82}:
\be
\langle \chi^2 \rangle = {H^3 t \over 2 \pi} \rar 
{\rm min} \left\{ (\sim H^4 /m^2),\,\, (\sim H^2 /\sqrt \lambda)
\right\}
\label{risechi}
\ee
Hence the field $\chi$ might acquire a large expectation value along
flat directions and when inflation is over and a small slope of the 
potential created by $m^2|\chi|^2$ becomes essential, the field returns
to the origin releasing baryonic (or leptonic) charge by the decay into 
quarks or leptons. However, it is important to keep in mind that if the 
field evolved only in radial direction so that $\dot \theta = 0$, the
baryonic charge accumulated by quantum fluctuations would be zero, as 
is seen from eq.~(\ref{bt}). Hence a motion in orthogonal direction is 
necessary for efficient baryogenesis. This motion may be produced by 
quantum fluctuations across the valley. So when the field $\chi$ 
``lives'' in 
the valley, far from the origin, it would oscillate between steep walls 
of the valley slowly approaching zero. When the field comes sufficiently 
close to the origin the quartic terms in the potential become negligible 
and the field
evolution is governed by the quadratic term $m^2 |\chi|^2$. In this 
region the field rotates around the origin with a decreasing amplitude. 
The damping of oscillation both in the valleys and near 
origin is forced by the following two phenomena: by the Hubble 
friction that diminishes the amplitude according to $|\chi| \sim a^{-3/2}$
and by the particle production which is strongly model dependent and might
be very much different for baryonic and leptonic $\chi$'s. Estimates of
the particle production rate can be found in ref.~\cite{dolgov92}. This 
effect may strongly damp motion along orthogonal direction in the valleys
and significantly diminish the generation of the charge asymmetry. 
The decay of $\chi$ at this stage creates average zero baryonic charge 
because the angular velocity $\dot\theta$ frequently changes sign. 
On the other hand, the 
decay of the field on the final stage, when it rotates around the origin,
proceeded with charge conservation and the produced quarks or leptons 
would have baryonic or leptonic charge equal to the initial magnitude of
the angular momentum of $\chi$, at the moment when $\chi$ 
arrived to the flat region and oscillations in angular direction changed 
into rotation. The magnitude of the latter is determined by the damping 
in the valley, while direction of rotation (i.e. the sign of baryon or
lepton asymmetry) is determined stochastically by the direction of
quantum fluctuations across the valley during inflation. In such a model
C and CP non-conservation is not necessary. The latter is created by
chaotic initial conditions. The size of domains with a certain
sign of the charge asymmetry may be very large due to exponential 
expansion. However, a concrete realization of the model can meet
serious problems and some fine-tuning of parameters or initial 
conditions may be necessary.

The asymmetry generated by the decay of $\chi$ at the final stage 
could be very large if the life-time of $\chi$ with respect to decay 
(or to the
particle production rate) is sufficiently long. The oscillating massive
field has the non-relativistic equation of state, $P=0$ (here $P$ is the
pressure density), so the energy density of such field decreases as
$1/a^3$. It is one power of $a$ slower than the decrease of the energy 
density of relativistic matter. The slower red-shift may create the 
situation when the matter in the universe is completely dominated 
by baryons (or leptons) with a negligible amount of antiparticles.

If C (CP) violation is explicit then the picture may be somewhat 
different. In our toy model we can realize this case assuming that
coupling constant $\lambda$ and mass are complex, so that the potential
takes the form:
\be
U(\chi) = m^2 \chi^2 + (m^*)^2 (\chi^*)^2 +\lambda \chi^4 + 
\lambda^*(\chi^*)^4
\label{complu}
\ee
Both terms are necessary, otherwise the CP-odd phase can be rotated 
away. In this potential the field $\chi (t)$ may acquire a non-vanishing
angular momentum after it traveled down along the bottom of the valley
without any orthogonal motion and closer to the origin it changes
``rails'' from the lambda-valley to the mass-valley that has a 
different direction. The transition from the bottom of one valley to 
another creates a transverse motion or, in other words, a 
nonzero baryonic or leptonic charge.

\section{Large and Inhomogeneous Lepton Asymmetry \label{lar}}

Models based on the picture described in the previous section
allow to obtain a small baryonic and a large leptonic asymmetry if 
e.g. the decay of the baryonic $\chi$-field is faster than the decay 
of the leptonic one. In this case baryonic angular fluctuations in 
the valleys would be stronger suppressed than leptonic ones, so the 
initial value of the baryonic charge at the beginning of rotation 
would be smaller than the initial value of the leptonic charge. 
Moreover, the dilution of the produced baryons by expansion could 
be stronger than the dilution of later created leptons. 

The characteristic wave length of the variation of the baryon 
($l_B$) or lepton ($l_L$) charge asymmetry 
was estimated in ref.~\cite{dolgov92} and is approximately equal to the 
smaller of $H_i^{-1} \exp( 1/\sqrt \lambda) $ and 
$H_i^{-1} \exp[(H_i/m)^2]$. In particular, it is possible to create
the universe with a small and homogeneous baryon asymmetry and large (of
the order of unity) lepton asymmetry, which in addition may be 
inhomogeneous at the scales well inside the present day 
horizon~\cite{dolgov91,dolgov99,dolgov98}. Such domains of leptonic charge
would create very strong density inhomogeneities and, in turn, too large
angular fluctuations of CMBR. However the flavor symmetry between
different lepton families, $e\lrar \mu\lrar \tau$, permits to avoid 
this problem because in this case one would expect domains that are
symmetric with respect to permutation of different leptonic charges  
and in first approximation the density contrast is 
vanishing~\cite{dolgov99,dolgov98}. 

In the simplest version of this scenario the universe would be filled
with equally probable domains with $(L_e,L_\mu,L_\tau) = (\pm 1,0,0)$
and with all possible permutations between leptonic charges. Evidently 
there are no walls between such domains, as follows from the 
way of their creation, and so the cosmological problem of heavy domain 
walls~\cite{zeldovich74} is avoided. Since the abundances of light 
elements produced at big bang nucleosynthesis are especially sensitive
to charge asymmetry of electronic neutrinos, there would be two types
of anomalous domains with low and high abundances and 4 types of the
normal ones with the same normal abundances, as is written in the
introduction. 

As discussed in the papers~\cite{dolgov99,dolgov98} the characteristic 
distances between the domains should be larger than a few hundred Mpc 
to agree with the observed small angular fluctuations of the CMBR 
temperature, $\delta T /T$. Another possibility of small size domains, 
that could create a large $\delta T /T$ on very small scales, which 
are allowed by observations, was not explored but possibly, though 
not surely, this option is excluded by direct observation of the light 
element abundances. Still it remains surprising that twice larger mass 
fraction of $He^4$ at not so large distances is not yet observationally 
excluded. Another signature of such inhomogeneities in primordial 
$He^4$ is a strong variation of the exponential slope in $\delta T /T$
at small angles (at $l >10^3$) at different patches on the sky. 
Possibly it is the most promising way to obtain an upper limit on 
inhomogeneities of primordial $He^4$.

\section{Inflaton Induced Phase Transitions and 
Baryonic Islands \label{inf}}

If the field $\phi$ that is responsible for baryo(lepto)-genesis is 
coupled to inflaton then a very interesting pattern of charge asymmetries 
may be created. This field $\phi$ could be e.g. the Affleck-Dine 
one described in the previous section or the complex scalar field whose 
condensate creates spontaneous breaking of charge symmetry~\cite{lee73}. 
The coupling to inflaton, $\Phi$, is assumed to be of the general
renormalizable form:
\be
{\cal L}_{int} = \lambda |\phi |^2 \Phi^2 + g |\phi|^2 \Phi =
\lambda (\Phi - \Phi_1)^2 |\phi|^2 - \lambda \Phi_1^2 |\phi|^2
\label{lphi}
\ee
where $\Phi_1 = g/2\lambda $ is a constant. We assumed that the 
interaction conserves C (CP), though it is not necessary. 

Generically the effective mass of the field $\phi$ may contain the
following contributions:
\be
\left( m_{eff}^{\phi}\right)^2 = m_0^2 + \xi R + \beta T^2 + 
\lambda (\Phi -\Phi_1)^2
\label{mphi}
\ee
where $R$ is the curvature scalar and the correspoding term comes from
a possible non-minimal coupling to gravity. The third term comes from 
the interaction with thermal bath, and the last one comes from the 
interaction with inflaton~(\ref{lphi}). Note that at inflationary stage
$R=12 H^2$, while at radiation dominated stage $R=0$. With such a form
of mass term one can easily visualize the following picture. At an early
inflationary stage the effective mass squared of $\phi$ is large and
positive, so that $\phi$ sits near the origin $\phi = 0$. When the 
inflaton evolves down and approaches the value $\Phi \approx \Phi_1$
the mass squared becomes negative and remains such for some period of
time determined by the inflaton evolution. When the amplitude of the
inflaton tends to zero and inflation ends, $m^2_{eff}$ may become 
positive again. During a relatively short period of negative $m^2$ the
road is open for $\phi$ to travel away from the origin and its 
subsequent destiny depends upon its self-interaction potential. The
field may either return to the origin, or stay for a while in a false
vacuum state with $\langle \phi \rangle \neq 0$, or evolve far away along 
one or other flat directions of the potential. Correspondingly the
generation of baryon (or lepton) asymmetry would be quite different. 
Some striking possibilities are discussed in the
papers~\cite{dolgov86,dolgov87,dolgov93} or in the review~\cite{dolgov92}.

In particular, our universe may be a huge baryonic island floating in 
the sea of dark matter~\cite{dolgov86,dolgov87}. This can be achieved if
the field $\phi$ is responsible for the spontaneous breaking of C and CP
symmetries~\cite{lee73}, so that cosmological charge asymmetry might be 
generated only in the regions where $\langle \phi \rangle \neq 0$. The 
size of this island may be of the order of the present day horizon or
much larger. In the first case a small observed dipole asymmetry of 
CMBR forces us to live very close to the center of the island and it makes
the model rather unnatural. Its interesting property is that the angular 
distribution of the CMBR temperature in this model would have an intrinsic 
dipole moment:
\be
D = \left({\delta T \over T} \right)_{dipole} \sim \left({ d\over R_i}
\right)
\label{dip}
\ee
where $R_i$ is the radius of the island and $d$ is our displacement from
the Center of the world. In this case the dipole contribution into 
angular fluctuations of CMBR is partly given by our peculiar motion and 
partly by the displacement from the Center and one does not need the very
large value of the velocity, about 600 km/sec, whose origin is mysterious,
to explain the observed dipole. There should also be an intrinsic 
quadrupole with a rather large value $Q\sim (d/ R_i)^2$, which is close 
to the observed one. 

The size of the island may be of super-horizon scale but nevertheless 
the angular fluctuations imprinted on CMBR by the existence of the 
boundary could be observable~\cite{grishchuk78}.
Of course their magnitude would be much 
smaller than in the sub-horizon case and we may live far from the Center
without breaking the observational limits. Still the large magnitude of
the dipole asymmetry of CMBR may be explained by the insular hypothesis.

Models of baryogenesis with spontaneous breaking of charge symmetry 
predict formation of cosmological matter-antimatter domains, as was 
noticed in ref.~\cite{brown79}. The size of these domains may be 
cosmologically large if after their formation the universe passed 
through a period of exponential expansion (inflation)~\cite{sato81}. 
A review of the earlier 
ideas on the subject can be found in the paper~\cite{stecker85} or in 
the recent one~\cite{dolgov00}, which contains a discussion and the list 
of references to later development. In the model described above C(CP) 
is broken spontaneously but only in finite bubbles, while outside the
bubbles particles and antiparticles are completely symmetric and 
cosmological charge asymmetry in these regions is vanishing. As in any
model with spontaneous breaking of charge invariance the universe as a
whole should be charge symmetric, so there must be an equal number of 
bubbles and anti-bubbles. We assumed above that the size of the bubbles
(or at least of our bubble) is very large, close to the horizon or much 
larger than it. In such a model other bubbles most probably should be 
far beyond the horizon and be unobservable.

In principle another scenario is possible when the bubbles are 
relatively small so that there are many of them inside the horizon
and the universe reminds a piece of Swiss cheese where bubbles of
matter and antimatter are separated by voids filled with dark matter.
There is no direct contact between them and thus no 
$p\bar p$-annihilation. This weakens the limits on antimatter domains
found in refs.~\cite{steigman76,cohen98} but the observed distribution 
of baryonic (luminous) matter seems to disagree with such a picture.
Moreover, angular variation of the CMBR temperature induced by
such isocurvature perturbations are likely to be unacceptably large. 
A related analysis was performed in references~\cite{kinney97,cohen97} 
but the case considered there rather differs from this model and some 
more work is desirable. 

\section{Large Baryon Asymmetry, Primordial Black Holes, and Dark
Matter \label{pri}}

In the model of the previous section the magnitude of baryon
asymmetry inside the islands was normal (\ref{beta}), but the sign 
could be different. The sea outside the islands was charge symmetric.
Here we will consider another exotic scenario of baryogenesis 
according to which the average baryon asymmetry is normal and 
positive but in its background there could be
relatively small (but still astronomically large) bubbles with a very
large (even close to unity) asymmetry. Such a picture can
be realized with the help of the scalar baryon discussed in 
sec.~\ref{bar} if the flat directions of its potential are ``locked''
by a positive value of $m^2_{eff}$ during almost all inflationary stage
and the window is only open when the inflaton field $\Phi$ is close to
$\Phi_1$, see eq.~(\ref{mphi}). In such a scenario the Affleck-Dine
field $\chi $ may reach large values and accumulate a big baryonic charge
only in relatively small bubbles, where the field succeeded to reach a
certain critical value. This process is stochastic and its probability 
may be sufficiently small so only a minor fraction of the universe
would be occupied by this baryon rich bubbles (see 
ref.~\cite{dolgov93} for the details). If the potential of $\phi$
is C(CP)-symmetric then the probability of baryon rich and 
antibaryon rich bubbles would be the same. So according to this scenario
the cosmological background is baryo-asymmetric and each bubble is
(strongly) baryo-asymmetric but on the average there is equal amount
of baryonic and anti-baryonic bubbles.

The subsequent destiny of these baryon rich bubbles depends upon their
size and the value of the baryon asymmetry. Predominantly they would
end as primordial black holes formed at the stage when baryons became
nonrelativistic. According to calculations of ref.~\cite{dolgov93} the
mass spectrum of the black holes in a simple version of the model is 
given by the log-normal distribution:
\be
{dN \over dM} = C \exp\left( -\gamma \ln^2 {M \over M_1} \right)
\label{dndm}
\ee
where $C$, $\gamma$, and $M_1$ are constants expressed through some 
unknown parameters of the underlying theory. We cannot make any 
reliable prediction about their value. They may easily vary by several
orders of magnitude. So in what follows we will treat them as free 
parameters to be constrained from observations.

According to this model not only the bubble sizes are stochastically 
distributed but also the magnitude of baryon asymmetry inside them. 
Depending upon the values of these two parameters the bubbles would
either form black holes and whether they are baryonic or anti-baryonic
could not be distinguished, because black holes do not have baryonic
hairs. Smaller size objects or those with a smaller value of the
asymmetry do not collapse and on the tail of the distribution we may
expect to observe clouds of antimatter in close vicinity, anti-stars,
or even possibly small anti-galaxies. 

Another interesting feature of this baryon (or antibaryon) rich regions
is that their primordial chemistry should be very much different from 
the standard one. The outcome of light elements produced during 
``the first three minutes'' strongly depends upon the baryon-to-photon
ratio, $N_B /N_\gamma$. Normally it is very small, eq.~(\ref{beta}),
and the nucleosynthesis effectively stops at 
$He^4$. The amount of the next stable light element $Li$ is roughly
9 orders of magnitude smaller. In the case of a larger ratio
$N_B /N_\gamma$ the amount of heavier elements could be much larger
and the baryon rich bubbles might have much more evolved chemistry which 
is more typical for the system contaminated by stellar processes.

The described here mechanism easily explains an early production of 
quasars, while in the standard theory of structure formation this
creates a serious problem. The model also explains an evolved 
chemistry around quasars. This could be the matter from baryon rich
region that did not completely disappeared in the black hole (quasar)
and which was enriched by heavier elements from the ``very beginning'',
i.e. from big bang nucleosynthesis. In the standard approach it is 
assumed that heavier elements observed in the vicinity of quasars were 
produced as a result of stellar processes in the stars of first 
generation. But this is very difficult to accomodate to the standard
theory of structure formation according to which structure are formed
much later than the observed distant quasars.

An interesting implication of the model of primordial black hole 
formation with the mass spectrum~(\ref{dndm}) is that they might
make a new form of cold dark matter. In contrast to the usual type
of dark matter, when all dark matter particles have equal masses, this 
model predicts a mass dispersed CDM where particles of different masses
play different role. The parameters describing the spectrum can
be fixed from the observations. In particular, one could request that 
there is one 
super-heavy black hole ($M>10^6 M_{\odot}$) per large galaxy. Here
$M_{\odot}$ is the solar mass. These super-heavy black holes can serve 
as seeds for galaxy formation. The bulk of dark matter may be presented 
by black holes with the masses near $10-100\,M_{\odot}$, while those 
with the mass about $0.5\,M_{\odot}$ may explain the MACHO's 
observations~\cite{milsztajn00,alcock00}. In addition the model gives
an explanation of formation of heavy ($M> 100 M_{\odot}$) black holes
observed in galaxies. A possible simple fit: 
$\gamma = 0.14$ and $M_1 = 0.5 M_{\odot}$, gives for the
relative contribution to the universe mass density in different mass
intervals the following results: $F(M<0.1) = 0.3\%$,
$F(0.1-2) = 12\%$, $F(M<10) = 38\%$, $F(M=2-30) = 48.5\%$,
$F(M=30-1000) = 37.5\%$, $F(M=10^3-10^6) = 1.6\%$,
$F(M=10^6-10^8) = 3.6\cdot 10^{-9}$, where $M$ is expressed in solar 
mass units. Of course a more accurate analysis with the account of 
existing astronomical bounds on the cosmic density of black holes in
different mass intervals is necessary to make more definitive 
conclusion~\cite{dolgov01} and numerical simulation of
structure formation in CDM model with the mass spectrum of dark matter
particles given by eq.~(\ref{dndm}) is desirable.

%
%

\end{document}